\documentclass[reprint]{revtex4-2}
\usepackage{graphicx} 
\usepackage{amsmath}
\usepackage{natbib}

\begin{document} 
\title{Integrated emitters with CMOS-compatible tuning for large scale quantum SiN photonic circuits}
\author{Jasper De Witte$^{1,2}$, Atefeh Shadmani$^{3}$, Zhe Liu$^{3,4}$, Andraz Debevc$^{5}$, Tom Vandekerckhove$^{1,2}$,\\ Marcus Albrechtsen$^{3}$, Rüdiger Schott$^{6}$, Arne Ludwig$^{6}$, Janez Krč$^{5}$, Gunther Roelkens$^{1,2}$,\\ Leonardo Midolo$^{3}$, Bart Kuyken$^{1,2}$ and Dries Van Thourhout$^{1,2}$}

\affiliation{$^1$Photonics Research Group, Ghent University-imec, Ghent, Belgium\\%
    $^2$Center for Nano- and Biophotonics (NB-Photonics), Ghent University, Ghent, Belgium\\%
    $^3$Center for Hybrid Quantum Networks, Niels Bohr Institute, University of Copenhagen, Copenhagen, Denmark\\%
    $^4$Sparrow Quantum ApS, Copenhagen, Denmark\\
    $^5$Faculty of Electrical Engineering, University of Ljubljana, Ljubljana, Slovenia  \\
    $^{6}$ Experimental Physics VI, Faculty of Physics and Astronomy,  Ruhr University Bochum, Bochum, Germany}

\maketitle

\section{Abstract}
Next-generation scalable quantum photonic technologies operating at the single photon level rely on bringing together optimized quantum building blocks with minimal optical coupling losses. Achieving this necessitates the heterogeneous integration of different elements onto a single interposer chip. Integrated quantum emitters are key enablers for generating single photons, inducing quantum nonlinearities, and producing entanglement. In this work, we demonstrate the scalable integration of mature InGaAs quantum dots embedded in GaAs waveguides onto a low-loss SiN photonic platform, as evidenced by a high processing yield of $94.7 \%$ using a commercially available micro-transfer printing tool. These integrated emitters are embedded within a p-i-n heterostructure that allows for noise suppression, near-blinking-free operation and wavelength tunability upon CMOS-level electrical biasing. With this, we pave the way for scalable integration of diverse quantum photonic devices on a single chip.

\section{Introduction}

A reliable quantum emitter, generating photons as mobile carriers of quantum 
information, is a fundamental building block for the future quantum internet. It is a key element in many proposals for future quantum hardware, addressing applications in quantum communication and quantum computing.  The requirements imposed by such proposals on the quantum emitter are typically very stringent but devices based on In(Ga)As quantum dots embedded in GaAs waveguides, are close to reaching the necessary performance: they now routinely achieve on-demand photons with $99\%$ purity and a pairwise indistinguishability of $96\%$\cite{Uppu2020,chan2025practical}.

To reach this level of quantum coherence, the noise processes needed to be minimized. Especially charge noise plays a very important role. It has shown to be fully eliminated with electrical control by embedding a quantum dot in a p-i-n diode heterostructure, as evidenced by blinking-free emission and near transform-limited linewidths\cite{Kuhlmann2013,Somaschi2016, Uppu2020}. The junction further allows electrical tuning of the emitter, useful to compensate for variations from inhomogeneous broadening and bring multiple quantum dots into resonance\cite{papon2023independent}.

With the incorporation of AlGaAs tunneling layers between the doped and intrinsic regions, these quantum confined Stark shifts can be made very large\cite{bennett2010giant}. %
The formation of these Coulomb barriers around the emitter further enables the quantum dot to be loaded with a single electron or hole, creating charged trion states\cite{Warburton2013} that can be controlled by adjusting the gate voltage. In this way, spin up/down states are addressable\cite{Uppu2021} through the Zeeman splitting of the electron states in the presence of an external magnetic field\cite{Hoegele2004}. 
This technique has already shown promise for the deterministic generation of multiphoton entangled states\cite{meng2024deterministic}, a crucial step toward fusion-based photonic quantum computing\cite{bartolucci2023fusion}. 

However, in such proposals the emitter is part of a much larger optical system. For their actual implementation, the requirements on total system efficiency, from generation to detection, are still too high. Recent results show that photon losses of only $8 \%$ can be tolerated\cite{chan2025tailoring,chan2025practical}. 
Other applications such as quantum repeaters\cite{borregaard2020one} and advanced quantum key distribution protocols\cite{Koodyski2020, gonzalez2024device, steffinlongo2024long} similarly require overall losses to be below 10$\%$.

Nowadays, quantum photonic experiments are typically implemented in a hybrid configuration where the different constituent devices are realized on different chips. 
E.g. in \cite{wang2023deterministic,pont2024high}, a separate source chip that provides highly coherent photons is interfaced to an ultralow-loss processing chip.
The interconnection of these different platforms inevitably requires multiple fiber-to-chip couplings and long photon travel distances, associated with high optical losses.
Because of this loss penalty, such modular system architectures lack the perspective to meet the typical application demands discussed above. 
A promising alternative is heterogeneous integration of the different required components on to a single interposer chip, as this would avoid the many fiber-chip interfaces typical for current demonstrations.
Among potential low-loss interposer platforms, SiN stands out due to its large transparency window and ultra-low transmission losses\cite{Chanana2022, buzaverov2024silicon}. 
However, reliable integration of a quantum emitter, which can be electrically controlled on this platform remained elusive so far. It would unlock unprecedented scalability as well as functionalities that remained hitherto inaccessible. That includes the co-integration with other important building blocks such as single photon detectors\cite{Ferrari2018, esmaeil2021superconducting,psiquantum2025manufacturable} and high-speed modulators\cite{niels2025demonstration, rahman2025high} for which waferscale processes have already been demonstrated by several groups.  

\begin{figure*}[] 
    \centering
    \includegraphics[width=\textwidth]{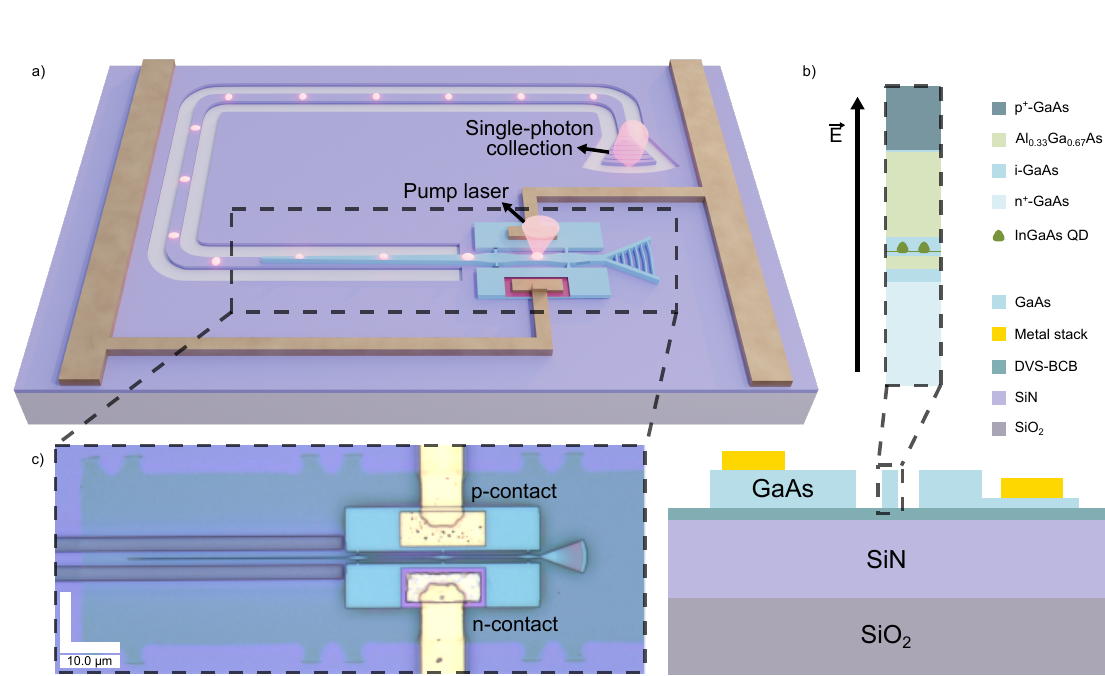}
    \caption{SiN/GaAs platform for quantum photonics a) Schematic of a micro-transfer printed GaAs nanobeam waveguide with diffraction grating coupler and tapered mode coupler to transmit light to the underlying SiN circuit. The pump laser beam used for top-excitation as well as the emitted single photons are indicated. b) Material layer stack of the printed coupon and SiN target. The metal layers on the p- and n-contacts are Cr/Au and Ni/Ge/Au stacks respectively while the final metallization on the SiN target as seen in part (a) of the figure is Ti/Au. Side tethers of the GaAs nanobeam for electrical biasing are not shown in this cross section. c) Microscope image of the same device from (a) after fabrication.}
    \label{fig:wow}
\end{figure*}

The first demonstrations of quantum dot emitters integrated onto SiN were based on die-to-wafer bonding\cite{Davanco2017, Shadmani2022}, a process that involves flipping the source layers onto the target substrate before removing the original substrate and etching the GaAs waveguides after bonding. While this is still an active line of research\cite{salamon2025electrical}, co-integration of the other important building blocks discussed above seems challenging. %
Alternatively, pick-and-place methods\cite{Mikulicz2024, Katsumi2020} allow for the integration of one functional device at a time. 
In a next step, it is desired to scale up such efforts towards high-throughput manufacturing. 
To that end, micro-transfer printing\cite{roelkens2024present}, is an interesting technique that combines the advantages of die-to-wafer bonding in terms of throughput with the flexibility of pick-and-place techniques.
It further allows for the prefabrication and pretesting of heterogeneous building blocks without interfering with the standard process flow of the interposer.
Regarding electrical control after heterogeneous integration, there is preliminary work using external electrodes\cite{Larocque2024} to create a capacitive structure over the III-V layer, InP in this reference. 
This approach requires very high voltages though and has been shown to remain vulnerable to screening effects from stray charges surrounding the quantum dot material. %
As established for emitters still on their native substrates, full suppression of charge noise demands the inclusion of a p-i-n heterostructure. In this work, we employ micro-transfer printing to integrate GaAs nanobeams that incorporate such a p-i-n structure. This enables active control of the emitter’s charge environment through Stark tuning with bias voltages below 0.6 V, well within the operating range of complementary metal-oxide-semiconductor (CMOS) technologies. %
We further demonstrate that the emitters retain their high purity and show no long-timescale blinking after transfer onto the SiN interposer.

To allow for future scaling to high-throughput integration as discussed above, we make use of a commercially available micro-transfer printing tool and achieve a high fabrication yield. Compared to custom-built tools that do not allow printing of many devices in parallel, these scalable tools are constrained in objective magnification and positional alignment, a limitation which is addressed in detail later. Still, the resulting high coupling efficiency from the GaAs to the SiN waveguides is shown to be preserved thanks to adaptations in design to accommodate for this misalignment.

\section{Results}

\begin{figure*}
    \centering
    \includegraphics[width=0.95\textwidth]{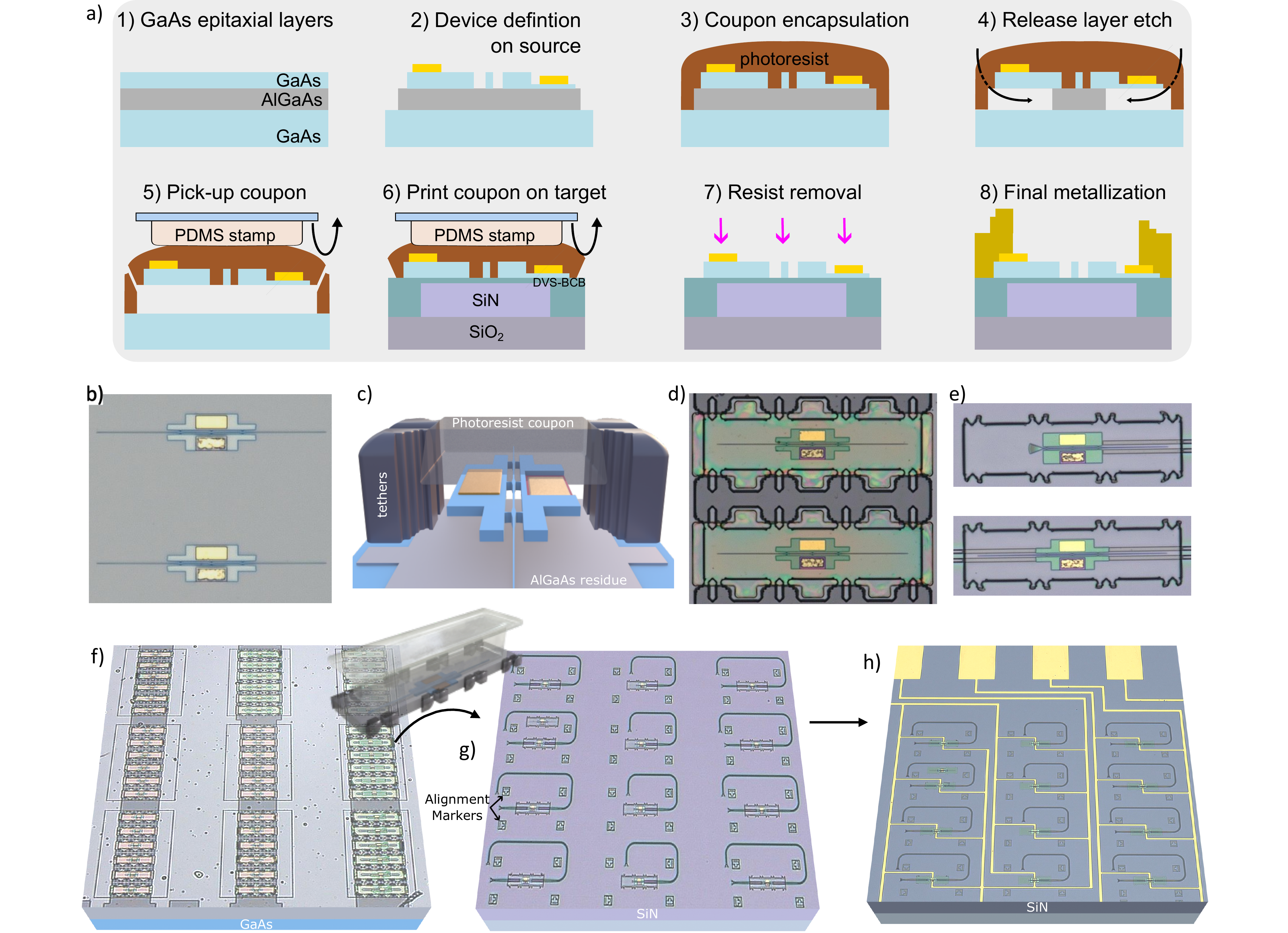}
    \caption{Fabrication process for heterogeneous integration a) Fabrication steps from predefinition on the GaAs source sample to micro-transfer printing and metallization on top of the target SiN sample. b) Microscope image of the GaAs devices predefined on the source sample c) 3D schematic cross section of the released coupon d) Microscope image of released coupons e) Printed coupons on the SiN interposer f-h) 3D impression from microscope images of a sample containing (f) released coupons ready for picking up, (g) printed coupons on the SiN interposer and (h) complete devices after photoresist removal and final metallization.
    }
    \label{fig:chip}
\end{figure*}

\subsection{Micro-transfer printing process}

Figure 1a and 1b show a schematic top view and cross-section of the proposed device.  It consists of a GaAs nanobeam integrated on a SiN waveguide.  The GaAs nanobeam has InGaAs quantum dots embedded in a vertical p-i-n structure. The detailed layer structure also includes the AlGaAs Coulomb barriers used for locking of the charge state, as discussed in the introduction.
The diode is electrically contacted through tethers at the side of the nanobeam visible in both the schematic top view and the microscope image in Fig. 1c. 

Figure 2 illustrates the process flow, with the different processing steps shown in Fig. 2a (we refer to the ``Methods'' section for processing details). 
The GaAs nanobeam sources, including their ohmic contacts, are first preprocessed  on their native substrate, using a well-tested all-soft-mask GaAs process flow\cite{Midolo2015}. The resulting structures are shown in Fig. 2b. 

In the common micro-transfer printing flow\cite{zhang2018transfer}, fully patterned photonic devices are tightly encapsulated within a dielectric or photoresist structure to form so-called coupons.
This then allows to create free-standing components by selectively etching an underlying release layer.
Such devices typically have dimensions of several micrometers making them mechanically very robust. In contrast, the GaAs nanobeams used here are only 160 nm thick and 300 nm wide in the emitter section to achieve a strong overlap of the optical mode with the quantum dots.
This makes them exceptionally fragile and prone to fracture during release or transfer. 
To mitigate this, each nanobeam is embedded within a larger rectangular photoresist coupon as can be seen in Fig. 2c-d.
These allow us to integrate different GaAs device geometries in the same way.
That includes nanobeam devices with different coupling structures at its input and output, such as grating couplers for fiber interfacing and tapered mode couplers for evanescent coupling to SiN waveguides.
To hold these coupons in place during subsequent release, the coupons include photoresist tethers, which are anchored to the GaAs substrate.

In the established GaAs process flow\cite{Midolo2015}, the underlying AlGaAs layer is removed in order to suspend the GaAs waveguides in air. 
Here, we repurpose this sacrificial layer to define free-standing photoresist coupons for micro-transfer printing, following a similar release strategy (Fig. 2a4). Figures 2c and 2d show a sketch and a microscope picture of the released coupons, respectively.

The target low-loss SiN sample (300 nm SiN/ 3.3 $\mu m$ SiO$_2$ / Si substrate) is prepared in parallel. In future work, this can be provided by a CMOS foundry. The SiN interposer contains waveguides with width of 880 nm and is coated with a 50 nm thick adhesive layer of divinylsiloxane bisbenzocyclobutene (DVS-BCB). Both samples are then loaded in the micro-transfer printing tool for the integration step. During transfer, an elastomer stamp made of poly-dimethylsiloxane (PDMS) is controlled with a piezoelectrically actuated stage and brought into contact with the free-standing coupon. Devices are picked-up one at the time by retracting the stamp, breaking the photoresist tethers in the process. They are then printed with automatic placement using pattern recognition on designated SiN markers for alignment. This way, we successfully integrated 38 quantum emitter devices onto the SiN platform, achieving a micro-transfer printing yield of $94.7 \%$ (36 successfully printed devices). After completing the printing procedure for all devices, the photoresist encapsulation is removed. A final metallization step is carried out to route the parallel electrical contacts across the chip (Fig. 2h). %

For the integration step, %
we employed a tabletop micro-transfer printer (X-Celeprint $\mu$TP-100). While single coupons were printed at the time, such tools can be used to scale this up to automated, high-throughput placement of arrays of devices in future work\cite{justice2012wafer}. However, the system introduces challenges not present in custom-built tools\cite{pholsen2025nanocavity} designed for single-device transfer. That is, the printer’s imaging objective has a limited 10x magnification, restricting the alignment precision typically below 0.5 micron for individual devices. This would remain to be the case when printing arrays of devices. To ensure efficient evanescent mode coupling between SiN and GaAs waveguides, the mentioned mode coupler design must tolerate these alignment variations without significant degradation in optical performance. As outlined in the ``Methods'' section and in Supplementary Note 3, we developed a three-step linear tapered mode coupler %
robust against realistic fabrication and alignment offsets. Variability in the exact optical mode transition —due to fabrication variations or positional inaccuracies during printing—was mitigated by adapting the GaAs tapering section to tolerate these shifts and widening the SiN waveguide to 3 $ \mu m$ in the coupling section. 

This offset in printed devices was characterized with a scanning electron microscope (SEM) (Fig. 3a). Figure 3b shows the design tolerance for both lateral and rotational misalignment. For those, we define the direction of misplacement as indicated on Fig. 3a. This makes the transmission plot of Fig. 3b asymmetrical in nature. The red crosses represent the experimentally measured lateral and rotational misalignment from the fabricated devices with a derived standard deviation of 152 nm and 0.085°, respectively, clearly situated in the region of $>99 \%$ transmission according to simulation. The impact of additional fabrication variations such as DVS-BCB thickness and waveguide width is presented in Supplementary Note 3. Overall, a high theoretical optical mode coupling efficiency $> 95 \%$ is expected within reasonable constraints imposed by the process.

\begin{figure}
    \centering
    \includegraphics[width=0.4\textwidth]{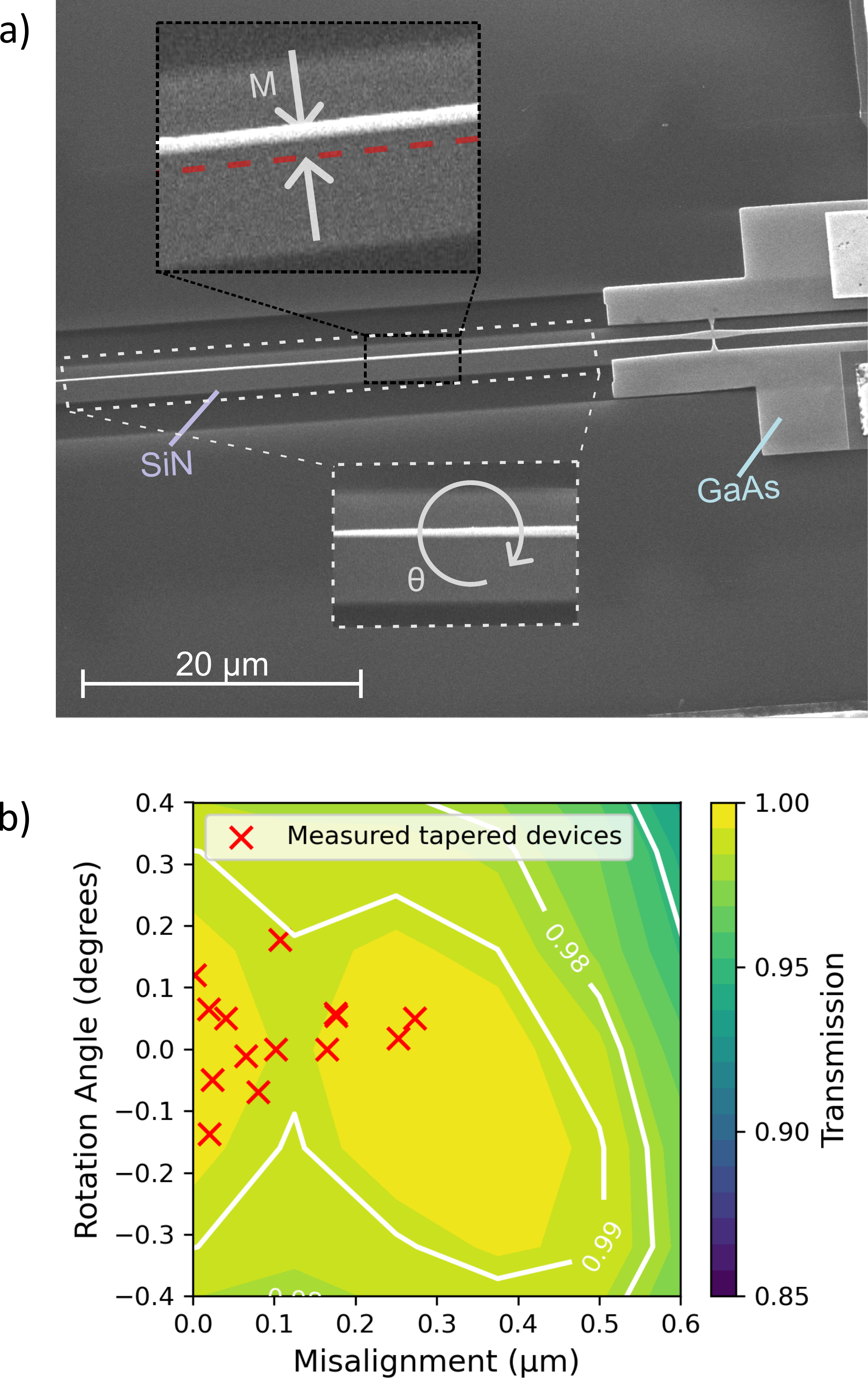}
    \caption{Assessment of micro-printing printing misalignment a) Scanning electron miroscope (SEM) image showing the GaAs nanobeam printed on top of the SiN target, indicating the lateral and possible rotational misalignment b) Contour plot of the simulation results for mode coupler transmission, performed with the eigenmode expansion method. 14 devices are experimentally characterized and their offsets are indicated with red crosses.
    }
    \label{fig:misalignment}
\end{figure}

\subsection{Optical and electrical properties}

\begin{figure*}[] 
    \centering
    \includegraphics[width=\textwidth]{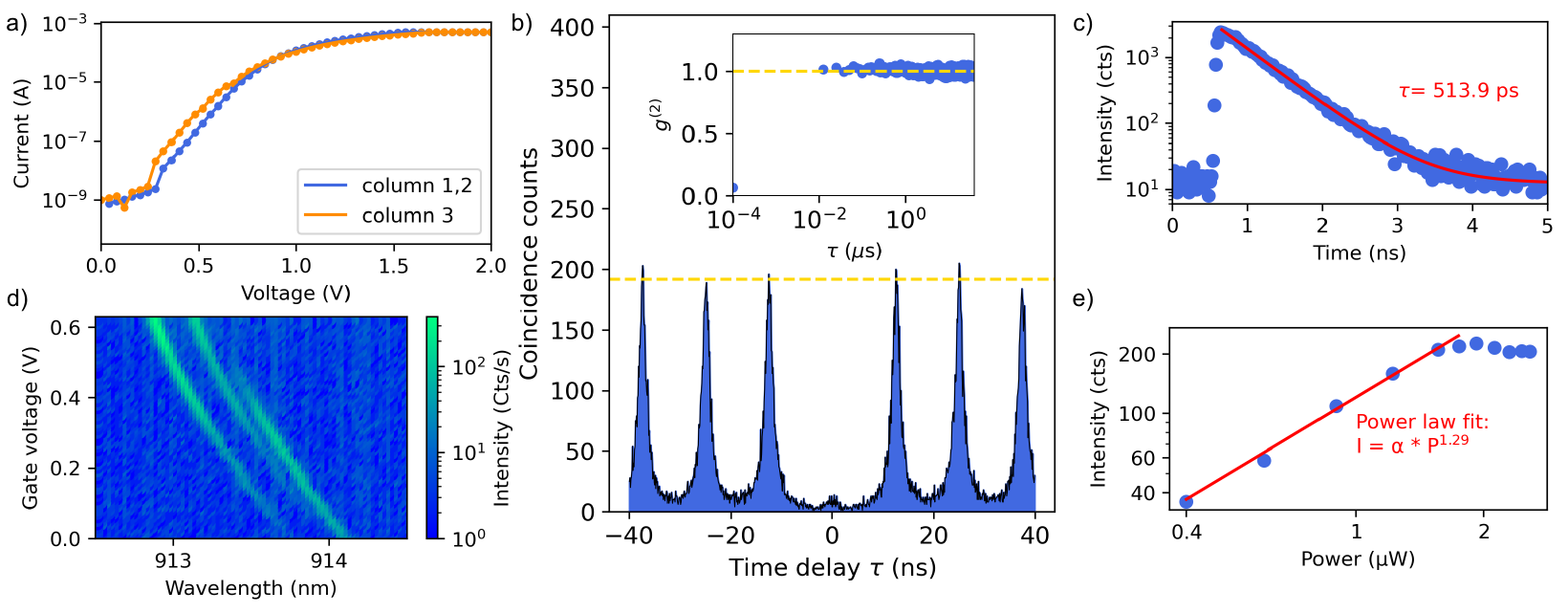}
    \caption{Quantum emitter properties of the heterogeneously integrated device a) IV curves of different columns of micro-transfer printed coupons. In addition to devices being contacted in parallel, columns 1 and 2 are also interconnected to each other. Relatively high currents are measured above 0.6 V b) Second order correlation function at short and long (inset) timescales. The horizontal yellow line indicates the average peak amplitude c) Lifetime measurement and single-exponential fit of the same emission line d) Photoluminescence spectrum as a function of bias voltage. The transitions are tuned in wavelength due to the quantum confined Stark effect e) Spectrometer counts as a function of pump power, fitted with an indicated power function.}
    \label{fig:tuning}
\end{figure*}
The sample was cooled to 4.3 Kelvin in a closed-cycle cryostat. With a built-in objective, light is coupled to on-chip grating couplers and the optical transmission of different device components was characterized at low temperatures with a continuous-wave laser at the target wavelength of 930 nm. The different device configurations allow us to systematically evaluate their loss contributions. SiN fully-etched grating couplers contribute to 11.5 dB of optical insertion losses, which can still be easily improved as discussed later on and in Supplementary Note 2. GaAs grating couplers exhibited similar performance, as they were not specifically adjusted for integration on top of a SiN chip. More relevant for this application is the performance of the mode coupler, designed to be insensitive to lateral printing misalignment as discussed above. For this, we can study the optical transmission measured through the devices from Fig. 2b: a GaAs emitter section with tapered mode couplers on each side. For this double-tapered GaAs device on top of SiN, we found a total transmission of -2.55 dB. Propagation losses in GaAs are estimated to be on the order of 21.25 dB/mm\cite{Wang2021} from previous results in the monolithic case.  In the 40 $\mu m$ long straight emitter section, this results in a loss of 0.85 dB. 
From this, we extract a remaining loss of another 0.85 dB ($82.2 \%$ transmission) per mode coupler. The discrepancy with the $> 95 \%$ efficiency predicted by simulation, can be attributed to remaining fabrication imperfections, as well as increased scattering losses from sidewall roughness in the narrow tapers. We examine this further in Supplementary Note 3. %
Before characterizing the quantum dot emission, the electrical IV curves of the parallel electrical contacts are inspected as shown in Fig. 4a. These exhibit clear diode behavior, demonstrating the reliable processing of the different metal contacts and connections. %
The relatively high current as well as its effect on emitter tuning are discussed in Supplementary Note 5.

\subsection{Quantum emitter performance}

Next, the quantum emitter properties were investigated. The objective inside the cryostat was now used to excite the quantum dots from the top as indicated in Fig. 1a and further detailed in the ``Methods'' section as well as in Supplementary Note 1. In the symmetrical GaAs waveguide, emission occurs bidirectionally and at most half of the photons can be collected from one side. As detailed in Supplementary Note 4, simulations predict a coupling efficiency of quantum dot emission with the fundamental GaAs nanobeam mode ($\beta$-factor) at either side of up to $90\%$, despite the extra underlying layers. 
Using an above-band excitation laser at 800 nm and scanning the applied bias below 0.6 V, we observed resolution-limited ($< 40$ pm) spectral lines from the quantum dots in the 905-920 nm range across all tested devices. %
By varying the bias voltage from 0 V to 0.6 V, we observed quadratic wavelength tuning of the quantum dot emission over a range of around 1 nm, as shown in Fig. 4d. 
Thanks to the thin Coulomb barrier near the n-contact\cite{Warburton2013}, %
no significant external bias is required to overcome the built-in voltage. As a result, quantum dot emission can be observed even at low bias. For more extensive photoluminescence spectra and emitter tuning plots over wider ranges of wavelength and bias voltage, we refer to Supplementary Note 5. These results are consistent with previous work on quantum dot wavelength tuning in monolithic devices from the same source wafer\cite{Gabi}. This shows that the micro-transfer printing process does not affect the wavelength tuning properties.

Under p-shell excitation, an isolated transition at 912 nm was selected using an optical filter with a full width at half maximum (FWHM) of 0.3 nm to remove phonon sidebands, other quantum dot lines, and stray pump light before the emission was sent to superconducting nanowire detectors. This setup enabled us to measure an exciton lifetime of 514 ps (Fig. 4c). Emission intensity increases with pump power according to the power law illustrated in Fig. 4e. To assess the purity of single-photon emission, we measured the second-order correlation function at excitation powers well below saturation. Figure 4b shows a strongly suppressed peak at zero delay. %
We extracted a g\textsuperscript{(2)}(0) value of $0.0572 \pm 0.0128$, confirming the single-photon nature of the emission. 
Analyzing the coincidence peak areas at long timescales (inset of Fig. 4b and Fig. 5), we observe low blinking below 2.75 $\%$ which can be associated with a radiative efficiency of 97.3 $\%$. This further confirms that the micro-transfer printing process did not alter the quantum dot decay dynamics nor did it introduce any detrimental defects acting as charge traps in the quantum dot surrounding. 

\section{Discussion}
Our micro-transfer printing method enhances passive low-loss SiN photonic integrated circuits that are foundry-compatible and mature with state-of-the-art coherent light-matter interfaces.  
On this platform, single photons can be guided through large circuits without suffering from the inevitable photon losses encountered in the native GaAs platform. This unlocks a wide range of novel applications in quantum information processing.
With the high fabrication yield achieved, future work should aim at co-integration\cite{soltanian2022micro, psiquantum2025manufacturable} with other quantum photonic components such as switches\cite{vanackere2023heterogeneous}, optical nonlinearities\cite{vandekerckhove2023high}, superconducting single-photon detectors\cite{tao2024single} and many more. %

Micro-transfer printing also enhances the GaAs material use thanks to the possibility of integrating emitter devices from a single source sample onto many SiN targets.
The reliability of our micro-transfer printing process can then be developed further similar to the general techniques outlined in \cite{roelkens2024present}. %
Specifically, %
the current fabrication yield can be improved by refining the tether breaking point and III-V processing. %
Currently, the transfer of a single device at the time takes around 3 minutes to complete. In comparison, further development of this process could lead to high throughput, automated processing of device arrays on reticle-sized stamps in printing cycles shorter than 60 s\cite{roelkens2022micro, niels2025demonstration}. 

For the interposer platform, our achieved GaAs-SiN coupling efficiency of $82.2 \%$ can be directly compared to the fiber-to-chip couplings required in the hybrid approaches discussed in the introduction. State-of-the-art SiN edge coupling already reaches near-unity efficiency\cite{psiquantum2025manufacturable}, shifting the main bottleneck to the outcoupling of the GaAs chip. Existing solutions based on shallow-etched grating couplers remain limited to efficiencies below $60 \%$\cite{zhou2018high, chan2025practical, meng2024deterministic}. Our platform is therefore expected to substantially outperform these hybrid strategies.

In the GaAs section, photon losses relate mostly to propagation losses. In addressing this, we do not foresee a possible reduction of the device length given the micro-transfer printing alignment constraints.
Propagation losses are primarily due to scattering from sidewall roughness\cite{payne1994theoretical} and electroabsorption\cite{Wang2021}. Since both effects are highly wavelength-dependent, a major improvement can be anticipated by transitioning to quantum dots that emit at higher wavelengths, such as the telecom bands, which are also compatible to low loss fibers\cite{holewa2024high,Larocque2024,ljubotina2025origins,albrechtsen2025quantum}. 

Future work might include more extensive spectroscopic experiments e.g. a study of the source coherence\cite{Uppu2020, Mikulicz2024} based on resonant excitation. That would benefit from an increased photon count rate. An important factor to the setup efficiency is a slight spectral mismatch between the designed grating coupler and the quantum dots used in this work. Besides this, coupling rates can be substantially improved by adopting advanced low-index grating coupler design\cite{lomonte2024scalable}, or by implementing horizontal fiber-to-chip coupling as discussed above.

Through Purcell enhancement, the photon generation rate could also be increased in optical cavities\cite{arcari2014near, Uppu2020}. This would further reduce the effects of charge and spin noise\cite{Kuhlmann2013} and improve nonlinear interaction strengths\cite{javadi2015single}. Finally, combined with the integrated AlGaAs Coulomb blockade and increased $\beta$-factor, this would then allow addressing of spin states in single trapped charges.

In conclusion, this work demonstrates a scalable and reliable micro-transfer printing procedure for the heterogeneous integration of quantum emitters onto a SiN platform using commercially available tools. The quantum emitter is embedded in a p-i-n diode heterostructure, which not only allows to tune the emitter transition frequency with CMOS compatible voltages but can also mitigate charge noise as evidenced by minimal emitter blinking. Our results pave the way for the co-integration of multiple cutting-edge quantum building blocks on a single low-loss interposer chip, thereby addressing the photon loss challenges of modular system architectures.

\section{Methods}
\subsection{Fabrication process}
Following the principles of \cite{Midolo2015}, the GaAs process relies on minimal plasma etching to protect the sensitive GaAs surfaces. The devices are patterned through electron-beam lithography (EBL) and inductively-coupled plasma etching (ICP). The n-type contact was formed from a Ni/Ge/Au stack, followed by rapid thermal annealing (RTA) at 420°C.  For the p-type contact, a Cr/Au stack was used. 
Photoresist coupons are defined with UV lithography. The anchoring of its tethers to the GaAs substrate is facilitated by creating openings through the sacrificial AlGaAs layer using ICP etching before the coupon encapsulation (Fig. 2a).
After a 1:30 ammonium dip to remove a possible native oxide on the AlGaAs layer, a 3:2 mixture of hydrogen chloride/distilled water is used at a temperature of 5 degrees Celcius to selectively remove this 1400 nm thick sacrificial layer and release the coupons. The chloride to water ratio is a trade-off between achieving an underetch that is complete and still gentle, not harming the free-standing photoresist coupons. 
In parallel, the SiN circuitry with a waveguide width of 880 nm is defined by EBL and RIE in a carbon tetrafluoride plasma. The target SiN interposer is coated with a 50 nm thick adhesive layer of divinylsiloxane bisbenzocyclobutene (DVS-BCB) which is then heated to 80 degrees to promote surface adhesion of the transferred coupon. Notably, this still limits the feasibility for array printing since heating the PDMS stamp adversely affects the configuration of its different posts. In the current process this is not a limitation as the PDMS stamp only contains a single post. The combination of alignment, picking and printing of each device takes around 3 minutes to complete, including a cleaning step of the PDMS stamp. The two failures were due to an incorrectly selected printing site and mechanical damage from a trapped particle underneath the device, respectively. We are confident this can be avoided in future work, since the process can be compared to micro-transfer printing of up to centimetre-long thin-film lithium niobate (TFLN) structures\cite{niels2025centimetre}, which is not limited by trapped particles or device cracks.

After printing, the photoresist encapsulation is removed with a combination of RIE processing in oxygen and sulfur hexafluoride plasmas combined with acetone rinsing. This way, also the DVS-BCB layer is almost completely removed apart from the coupon imprint which can be recognized by remaining green contour in Fig. 1c. Finally, the parallel electrical contacts are routed across the printed device columns to wirebonding pads at the side of the chip (cfr. Fig. 2h), with a metallization layer of Ti/Au.

\subsection{Mode coupler geometry}

The GaAs adiabatic mode coupler used to transmit light between the SiN and GaAs layer is designed similar to the optimization approach described for a stepwise linearly tapered coupler in \cite{shi2019novel}. The design is based on an analysis of the effective index and mode overlap of the SiN/GaAs supermode for different GaAs widths. The first taper section contains a narrow taper tip of 80 nm width that prevents reflections at the interface. This section is 5 $\mu m$ long and tapers quickly without exciting unwanted optical modes, up to a GaAs width of 115 nm where the hybrid mode begins coupling upwards into the GaAs layer. The following 35 $\mu m$ coupling section further tapers the width from 115 nm to 215 nm and guides the mode transition adiabatically into the GaAs. To account for fabrication imperfections, we have to allow some uncertainty on the exact point where maximal mode transition occurs. Therefore, this section is extended compared to the nominal design geometry. The last section tapering the GaAs width from 215 nm to 300 nm is again 5 $\mu m$ long to maintain an adiabatic mode transition. The underlying SiN waveguide is designed to be 3 $\mu m$ wide to reduce its sensitivity to printing misalignment.

\subsection{Experimental setup and optical loss budget}
Inside the cryostat, the chip is mounted on a printed circuit board (PCB).  Through wirebonding, electrical connections are made to the parallel contacted emitter device columns.  This PCB is further connected to a voltage supply outside of the cryostat.

The closed-cycle 4.3 K cryostat system further contains a built-in objective. With that, stable optical measurements could be maintained over the course of several hours. A detailed and graphical overview of this optical setup is provided in Supplementary Note 1. The objective is part of a 4f imaging system which allows excitation of a quantum dot from the top while collecting emitted photons through the grating couplers within its field of view. Input and output beam paths are separated in a 90/10 beam splitter, sending $90\%$ of the light into the collection path. Here, an ultra-narrowband filter (0.3 nm) is positioned in the infinity space of the setup to isolate a single optical transition before coupling the free-space beam back in an optical fiber.The signal is then guided through 60 m of fiber to superconducting nanowire detectors with $\sim 90 \%$ detection efficiency over the wavelength range of interest. 

\begin{table}
\centering
\begin{tabular}{ll}
Component            & Transmission  \\
                     & efficiency (\%) \\ \hline
unidirectional $\beta$-factor& 45 \\
GaAs/SiN coupling          & 82.2                        \\
SiN grating coupler      & 7.0                            \\
Collection objective & 90                           \\
Mirrors              & 90.2                         \\
90/10 beam splitter             & 90.0                           \\
Half-wave plate, & \\
polarization beam splitter            & 98.0                           \\
Fiber coupling       & 36.0                           \\
Fiber towards detector    & 26.1                         \\ 
SNSPD                & 90                           \\ \hline
Total                & 0.2                      
\end{tabular}
\caption{Breakdown of the setup efficiency throughout its different components}
\end{table}
Table 1 shows the optical transmission throughout the measurement setup from the integrated quantum emitter to detector. In this analysis, we neglected the propagation losses of the straight emitter section in which the quantum dots are randomly located. Clearly, the losses are largely dominated by coupling components (grating couplers, collimators and fiber) throughout the setup.

\subsection{Second order correlation function analysis}
From the second-order correlation measurement shown in Fig. 4b, we derived a raw g\textsuperscript{(2)}(0) value of $0.0702 \pm 0.0129$ by comparing the integral of the center peak at zero time delay with the 60\textsuperscript{th} peak. This already confirms the single-photon nature of the emission. 
By fitting the data to a convolution of the instrument response function and a double-sided exponential, we derive the corrected g\textsuperscript{(2)}(0) value of $0.0572 \pm 0.0128$ discussed in the main text. 
\begin{figure}
    \centering
    \includegraphics[width=0.4\textwidth]{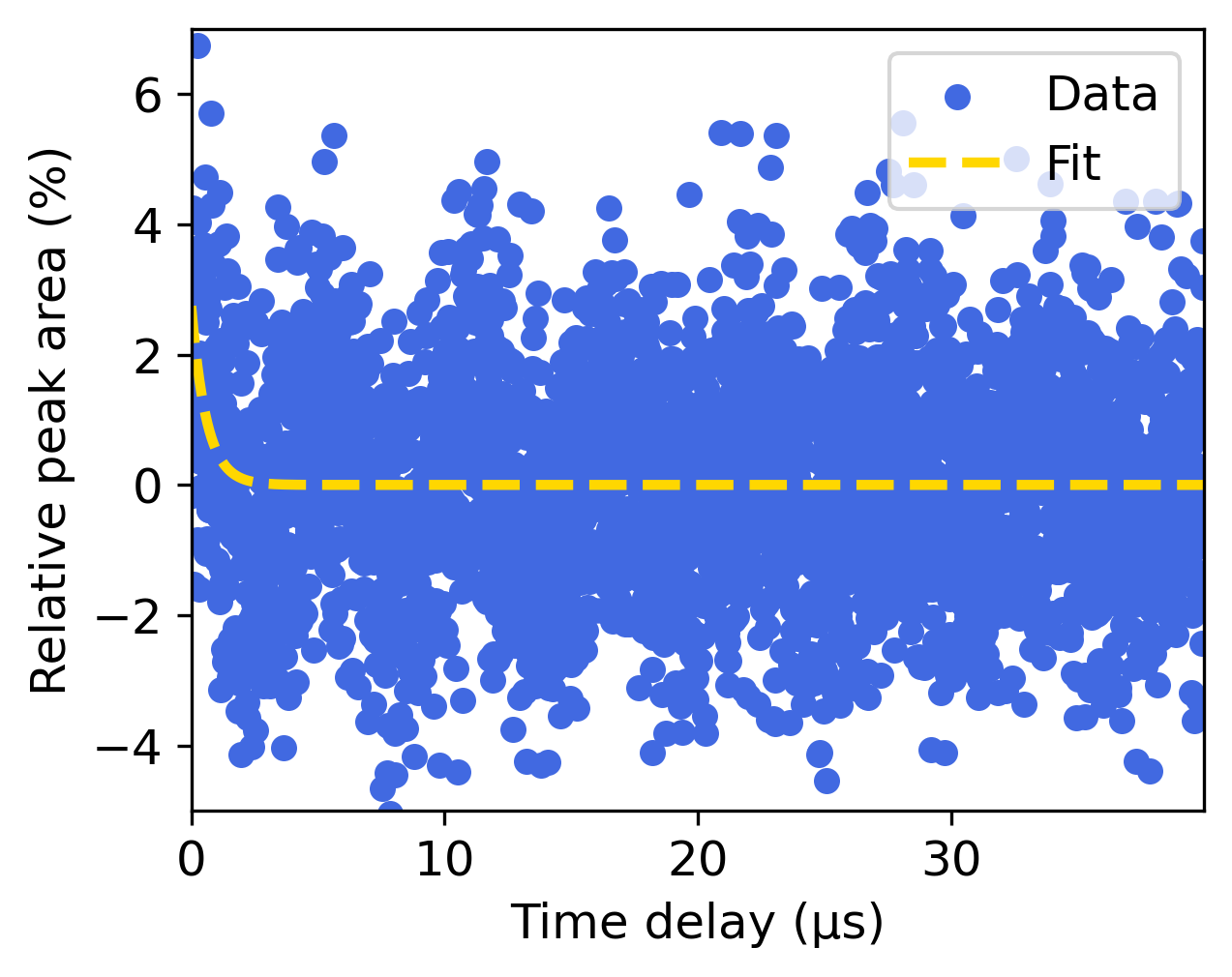}
    \caption{Relative area of side peaks from the correlation function at long positive time delays, showing low levels of emitter blinking. 
    }
    \label{fig:blinking}
\end{figure}
In Figure 5, we visualized the relative peak area from the inset of Fig. 4b in linear scale. This shows minimal blinking below 2.75 $\%$ as discussed in the main text.
At higher powers, we did observe some increased blinking behavior related to fluorescence dead times, indicated by bunching at short timescales. The quantum dot was therefore pumped at low power and the pump light was suppressed in the output part of the setup with the bandpass filter discussed above. Therefore, no background subtraction was required in this g\textsuperscript{(2)} analysis.

\subsection{Data availability}
The data that support the findings of this study are available from the corresponding author upon reasonable request. 

\section{References}

\section{Acknowledgements}
We thank Steven Verstuyft , Muhammad Muneeb and Liesbet Van Landschoot from Ghent University-imec for their invaluable aid in fabrication training and advice, Clemens Krückel from Ghent University-imec for his aid with the room temperature measurement setups. We finally thank Rodrigo A. Thomas from Center for Hybrid Quantum Networks for his aid with low temperature setups. We acknowledge the support by Fonds Wetenschappelijk Onderzoek (FWO) grants 1S69123N and 11H6723N, European Research Council (ERC) under the European Union’s Horizon 2020 research and innovation program (No. 949043, NANOMEQ) and the QuantERA II Programme with funding from the European Union’s Horizon 2020 research and innovation programme under Grant Agreement No 101017733. R.S. and A.L. acknowledge support by the BMFTR funded projects QTRAIN No. 13N17328, EQSOTIC No. 16KIS2061,and QR.N No. 16KIS2200, and the DFG funded project EXC ML4Q LU 2004/1

\section{Author contributions}

J.K, L.M., B.K. and D.V.T conceived the idea for the project.
J.D.W., A.S. and Z.L have developed the processes (coupon fabrication and transfer printing). R.S. and A.L. were responsible for material growth, J.D.W. and A.D. designed and numerically simulated the components. A.S. and M.A. built the cryostat setup. J.D.W. and A.S. performed initial measurements, A.S. fully characterized the system efficiency and gathered data on the final emitter properties. J.D.W., A.S. and T.V analyzed the results. J.D.W. and A.S. prepared figures, J.D.W. wrote the manuscript with input from coauthors. J.K, G.R., L.M., B.K. and D.V.T. supervised the project.

\section{Competing interests}
The authors declare no competing ﬁnancial interests.

\end{document}